\documentclass[conference]{IEEEtran} 

\usepackage[dvips]{graphicx}
\usepackage{subcaption}
\usepackage{epsf}
\usepackage{epsfig}
\usepackage[noadjust]{cite}
\usepackage{amsmath}
\usepackage{url}
\usepackage[export]{adjustbox}
\usepackage{soul}
\usepackage{algorithm}
\usepackage[noend]{algpseudocode}
\usepackage{color}
\usepackage{tree-dvips}
\usepackage{float}
\usepackage{multirow}
\usepackage{comment}
\usepackage{soul}
\usepackage{booktabs}
\usepackage[dvipsnames]{xcolor}
\usepackage{siunitx}
\usepackage{footnote}
\usepackage{threeparttable}

\def\BibTeX{{\rm B\kern-.05em{\sc i\kern-.025em b}\kern-.08em
    T\kern-.1667em\lower.7ex\hbox{E}\kern-.125emX}}

\newsavebox{\mybox}
\newlength{\mydepth}
\newlength{\myheight}

{\begin{lrbox}{\mybox}\begin{minipage}{\textwidth}}%
{\end{minipage}\end{lrbox}%
  \settodepth{\mydepth}{\usebox{\mybox}}%
  \settoheight{\myheight}{\usebox{\mybox}}%
  \addtolength{\myheight}{\mydepth}%
  \noindent\makebox[0pt]{\hspace{-20pt}\rule[-\mydepth]{1pt}{\myheight}}%
  \usebox{\mybox}}
  
\begin{document}

\onecolumn 

{\LARGE IEEE Copyright Notice} \\

\copyright 2019 IEEE. Personal use of this material is permitted. Permission from IEEE must be obtained for all other uses, in any current or future media, including reprinting/republishing this material for advertising or promotional purposes, creating
new collective works, for resale or redistribution to servers or lists, or reuse of any copyrighted component of this work in other works. \\

{\large Accepted to be Published in: Proceedings of the 2019 IEEE International Midwest Symposium on Circuits and Systems (MWSCAS), Aug. 4-7, 2019, Dallas, TX, USA.}

\twocolumn

\title{\vspace*{-1.5pt}An SR Flip-Flop based Physical Unclonable Functions for Hardware Security} 

\author{\IEEEauthorblockN{Rohith Prasad Challa, Sheikh Ariful Islam and Srinivas Katkoori } 
 \IEEEauthorblockA{Department of Computer Science and Engineering\\
 University of South Florida \\
 Tampa, FL 33620\\
 Email: \{challa1, sheikhariful, katkoori\}@mail.usf.edu}\vspace*{-0.99cm}}

\maketitle

\begin{abstract}

Physical Unclonable Functions (PUFs) have emerged
as a promising solution to identify and authenticate Integrated
Circuits (ICs). In this
paper, we propose a novel NAND-based Set-Reset (SR) Flip-flop
(FF) PUF design for security enclosures of the area- and power-constrained
Internet-of-Things (IoT) edge node. Such SR-FF based PUF is constructed
during a unique race condition that is (normally) avoided due to
inconsistency. We have shown, when both inputs (S and R) are
logic high (`1') and followed by logic zero (`0'), the outputs $Q$ and
$\overline{Q}$ can settle down to either 0 or 1 or vice-versa depending on
statistical delay variations in cross-coupled paths. We incorporate
the process variations during SPICE-level simulations to leverage the capability of SR-FF in
generating the unique identifier of an IC. Experimental results
for 90nm, 45nm, and 32nm process nodes show the robustness of SR-FF PUF responses in
terms of uniqueness, randomness, uniformity, and bit(s) biases.
Furthermore, we perform physical synthesis to evaluate  the applicability of SR FF
PUF on five designs from OpenCores in three design corners. The estimated overhead for power, timing, and area in three design corners are negligible.

\end{abstract}
\section{Introduction}

Due to the horizontal business model and vertical disintegration of IC design, most of ICs'  manufacturing and testing of fabless design houses are performed in foreign foundries. In the heart of this design ecosystem, original IP owners face several security challenges including overproduction, counterfeiting, authentication, and trust in manufactured products. Among all the possible existence of security solutions, Physical Unclonable Function (PUF) acts as one-way function that can map  certain stable inputs (challenges) to  pre-specified outputs (responses). Although cryptography algorithms have been put into practice to perform the authentication, they are difficult to upload due to recent attacks \cite{8031550,1561249}.  Moreover, the deployment of computationally intensive cryptographic algorithms in resource-constrained IoT devices limit their wide adoption. In contrast, PUF utilizes inherent silicon variations. If a similar design is manufactured onto two different dies,  process variations would act differently within and across  both dies and this forms the basis for a PUF. Ideally, a PUF implementation should be low-cost, tamper-evident, unclonable, and reproducible. The PUF response also need to be invariant to environmental variations

In recent years, a wide variety of PUF architectures have been investigated that can transform device properties (e.g. threshold voltage, temperature, gate length, oxide thickness, edge roughness) to 
a unique identifier of certain length.
Metastability in cross-coupled paths have been exploited to design PUF  with  SR latch \cite{HABIB201792,7881790,8351052} and Ring Oscillator (RO) \cite{BossuetNCF14}. Although latch-based PUF designs offer unique signatures to ICs, they suffer from signal skew and delay imbalance in signal routing paths and Error Correction Code (ECC) circuitry is commonly employed to post-process the instable PUF responses \cite{7428066}. On the contrary, RO-PUF in \cite{BossuetNCF14} incurs significant area overhead that includes a counter, an accumulator, and a shift register. These serve as a motivation to harvest deep-metastability in bi-stable memory, SR FF, to design a low-cost PUF and high-quality CRPs.

In this paper, we design and analyze a novel SR FF based PUF. For a NAND gate based SR FF, the input condition for S(Set) = `1' and R(Reset) = `1' must be avoided as it produces an inconsistent condition.  When S=R=`1' is applied followed by S=R=`0', the outputs $Q$ and
$\overline{Q}$ would undergo race condition. Due to manufacturing variations, the state due to race condition will settle in either `0' or `1'. Further, due to intra-chip process variations, some flip flops in a chip will settle in `0' state, while others will settle in `1' state.  In addition, due to inter-chip variations, such signature will be different across the chips. We investigate delay variations  in NAND gates of the feedback path  that affect most the gate delay. We validate the proposed idea with SPICE-level simulations for 90nm, 45nm, and 32nm process nodes to establish the robustness of the proposed PUF responses for  16-, 32-, 64-, and 128-bit responses. We also perform layout-level simulation with foundry data on five designs that incorporate SR-FF and present their figures of merit (power, timing, and area). In summary, we present the following novel contributions:

\begin{itemize}
\vspace{-1pt}

\item utilizes SR-FFs already present in the register of a design without any ECC and helper data. The responses are free from  multiple key establishments round that can thwart reliability based attack.

\item input dependent random yet stable binary sequence aided by unpredictable manufacturing variability. Depending on input challenges, only a fraction of SR-FFs would be utilized to create unique device signature. Therefore, it would increase the attacker reverse engineering effort to determine the exact location of such SR-FFs that participate in PUF responses generation. 

\item a centroid architecture such that surrounding transistor variations would only affect  PUF responses and evaluate the associated overhead through layout-based synthesis.
\end{itemize}

The rest of the paper is organized as follows: 
\mbox{Section \ref{sec:background}} provides background on the types of PUFs using metastability.
\mbox{Section \ref{sec:proposed work}} describes the construction of SR flip-flop based PUF design.
Section \ref{sec:experiments} reports in detail the experimental results. 
Finally, Section \ref{sec:conclusion} draws the conclusion and future work.

\section{Background and Related work}
\label{sec:background}

A  PUF is a digital fingerprint that serves as a unique identity to silicon ICs and characterized by  inter-chip and intra-chip variations. Inter-chip offers the uniqueness of a PUF that helps to conclude that the key produced for a die is different from other keys. Intra-chip determines the reliability of the key produced that should not change for multiple iterations on the same die. For a signal, metastability occurs when the specifications for setup and hold time are not met and unpredictable random value appears at the output. Although metastable is an unstable condition, due to process variations, such metastability generates a stable but random state (either `0' or `1'), which is not known {\em apriori}.

Transient Effect Ring Oscillator (TERO) PUF \cite{BossuetNCF14} utilizes metastability to generate the responses with  a binary counter, accumulator, and shift register. Although the architecture is scalable, it requires large hardware resources. Su {\em et al.} \cite{8351052, 8357321} presented cross-coupled logic gates  to create a digital ID based on threshold voltage. The architecture is composed of latch followed by a quantizer and a readout circuit to produce the PUF ID. However, readout circuit is generally expensive and limits its application to the low-power device. FPGA-based SR-latch PUF has been presented in \cite{7881790, HABIB201792}. Due to temporal operating conditions, ECC is employed to reliably map one-to-one challenge-response pair in both approaches. To alleviate power-up values from memory-based PUF, registers based on edge-triggered D-FF are proposed in \cite{8351052,8607170}. The authors suggested to include an expensive synchronizer in Clock Domain Signal (CDC) signals to get stable PUF response. A framework of `body-bias' adjusted voltage on SR-latch timing using FD-SOI technology is presented in \cite{8491861}. To get correct PUF response, authors employed buffers along the track of top and bottom of  latches that suffer from responses biasedness.

The majority of works utilizing metastability to design PUF employ additional hardware to count the oscillation frequency.  Our work is unlike these previous studies in that it
(a) employs SR-FF to construct low-cost PUF and (b) reuses the SR-FF already in the original IP by varying channel length and temperature to account for intra- and inter-chip variations.

\section{Proposed SR Flip Flop PUF}
\label{sec:proposed work}

\begin{figure}[h]
\vspace{-2ex}
\centering
\includegraphics[width=0.4\textwidth]{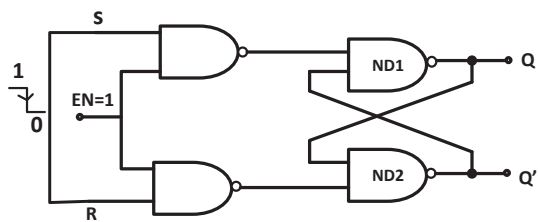}
\vspace{-2ex}
\caption{SR Flip Flop}
\label{fig:SR_FF}
\vspace{-5ex}
\end{figure}

Our approach presents a PUF design that relies on the cross-coupled path in an SR-FF configuration. Each bit of a PUF response can be extracted from metastability induced random value  in the output ($Q$) due to a particular input sequence at SR-FF. This random value would eventually evaluate to a stable logic due to process variability. A  clock enabled cross-coupled NAND-based SR-FF construction is shown in Fig. \ref{fig:SR_FF}. It does not require additional synchronizer to control the input conditions. Set-Reset (SR) Latch has the forbidden input combination, namely, S=R=1 which results in both $Q$ and $\overline{Q}$ equal to 1.  After S=R=1 input, if we lower both inputs (S=R=0), there is a race condition between the two cross-coupled NAND gates (ND1 and ND2) making $Q$ and $\overline{Q}$ to linger around $\frac{Vdd}{2}$ value. Although such race condition is prohibited during normal circuit operation, it can influence the output to generate a state determined by the mismatch in the underlying device parameters (transistor length, threshold voltage).  Analysis of the race behavior is seemingly dependent on precise phase relation between clock and input data. We exploit such input-referred event sequence to generate PUF response.

\begin{figure}[h]
\centering
\vspace{-1ex}
\includegraphics[width=\columnwidth]{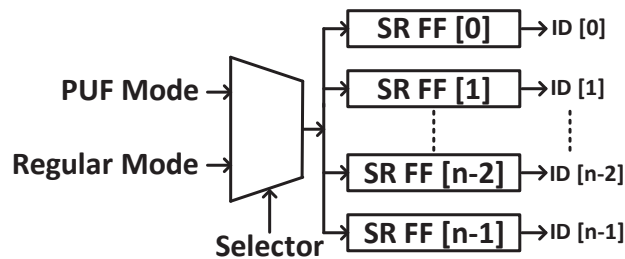}
\caption{Architecture of  dual-mode n-bit array SR FFs}
\label{fig:SRFF_chain}
\vspace{-2ex}
\end{figure}

Fig. \ref{fig:SRFF_chain} shows the architecture of n-bit array SR-FFs with an input multiplexer (MUX) to select either PUF or regular mode. As each SR-FF would generate a single bit key, we can obtain a PUF signature of the maximum size of FF instances. However, it suffers from multiplexer output that has to be sufficiently long to reach all SR-FF instances. It would also increase the delay to produce random output at $Q$ depending on the longest distance from MUX output to an SR-FF instance. As a result, both higher wire length from MUX output and longest transition time  due to metastability would decrease the timing performance of an SR-FF based PUF during regular operation. Furthermore, such architecture may be susceptible to key-guessing attack under a single clock pulse. Hence, the architecture in Fig. \ref{fig:SRFF_chain} would be biased towards variations in the connecting wire length and width. This, in turn, reduces the impact of transistors' local variation. In short, the higher the depth of PUF timing paths, the less its response would depend on transistors behavior.

\begin{figure}[h]
\centering
\vspace{-2ex}
\includegraphics[width=0.49\textwidth]{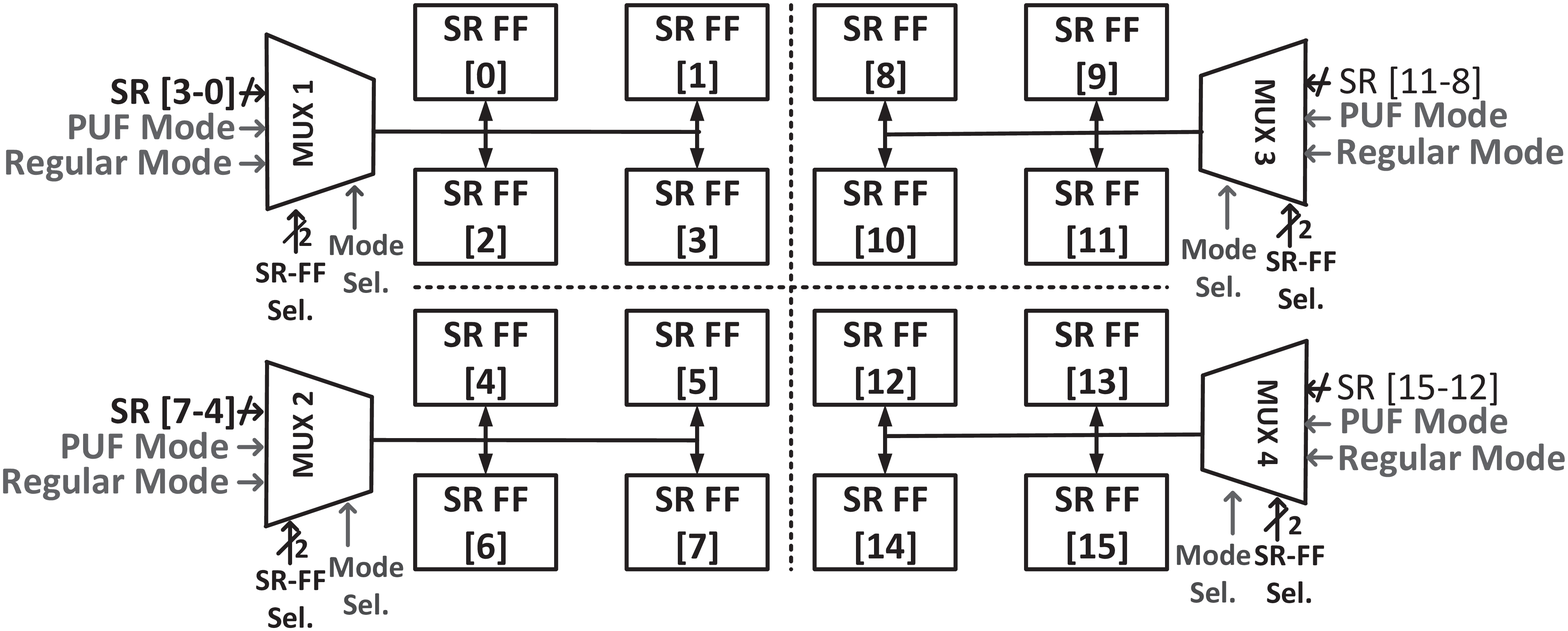}
\caption{Centroid of 16-bit SR Flip-Flops}
\label{fig:SRFF_centroid}
\vspace{-2ex}
\end{figure}

Fig. \ref{fig:SRFF_centroid} shows a centroid architecture (4x4 grid) of 16-bit SR-FFs built upon Fig. \ref{fig:SRFF_chain} with additional MUXs to improve the delay and thwarting the key-guessing  attack. It also results in improved bit distribution by preventing edge-effects \cite{4443209}. Each multiplexer has three selector bit, of which, two would be
used to select an SR-FF in a grid and the remaining would determine mode (PUF or normal) selection. A simple controller is embedded in the original architecture to aid in the signal extraction process.  Depending on the number of controllable MUXs, the size of the partitions can increase or decrease.

\section{Experimental Results}
\label{sec:experiments}


\begin{figure*}[h]
\centering
\resizebox{\textwidth}{!}{
\begin{tabular}{ccc}
\includegraphics[width=\textwidth]{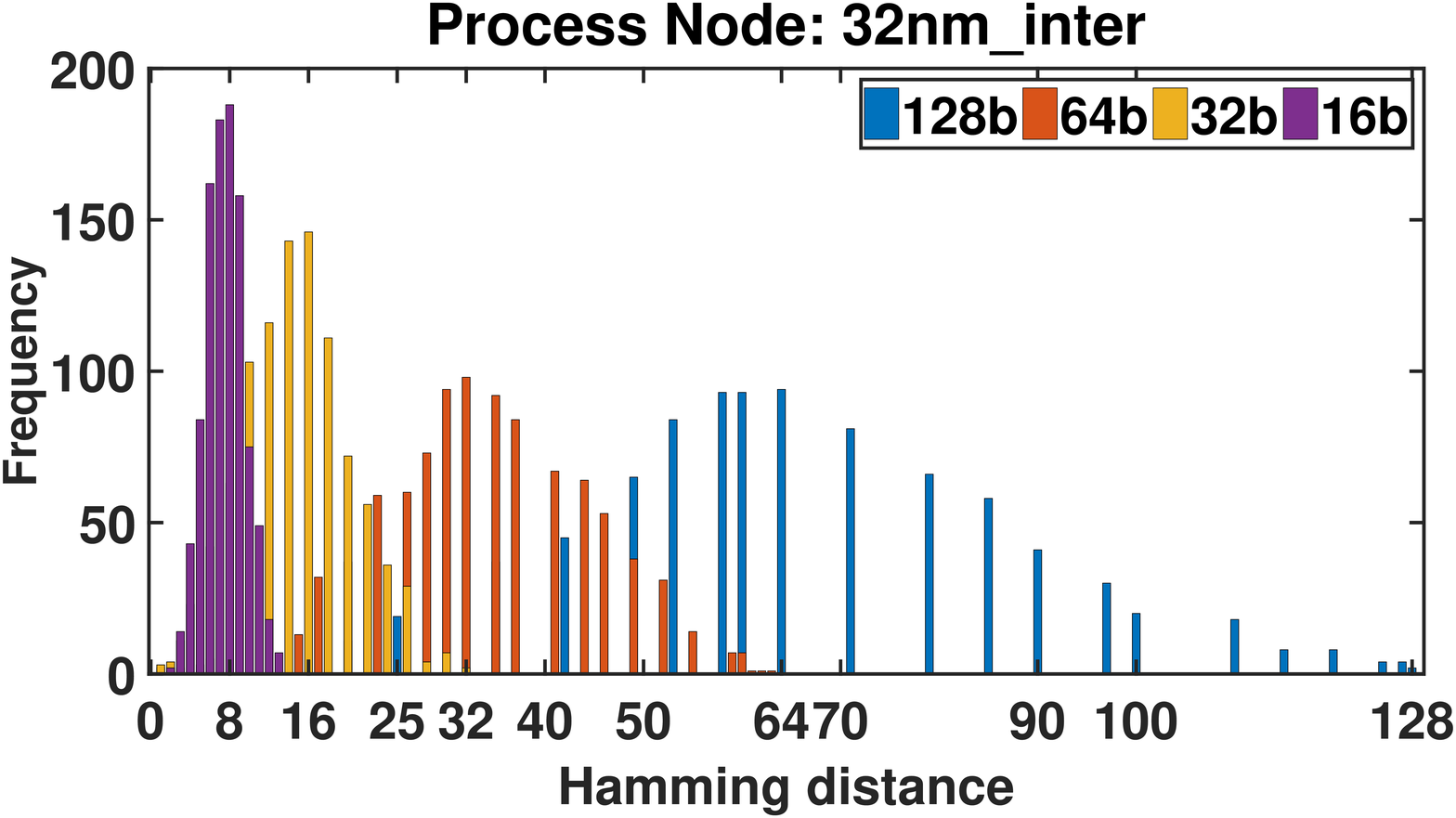} & 
\includegraphics[width=\textwidth]{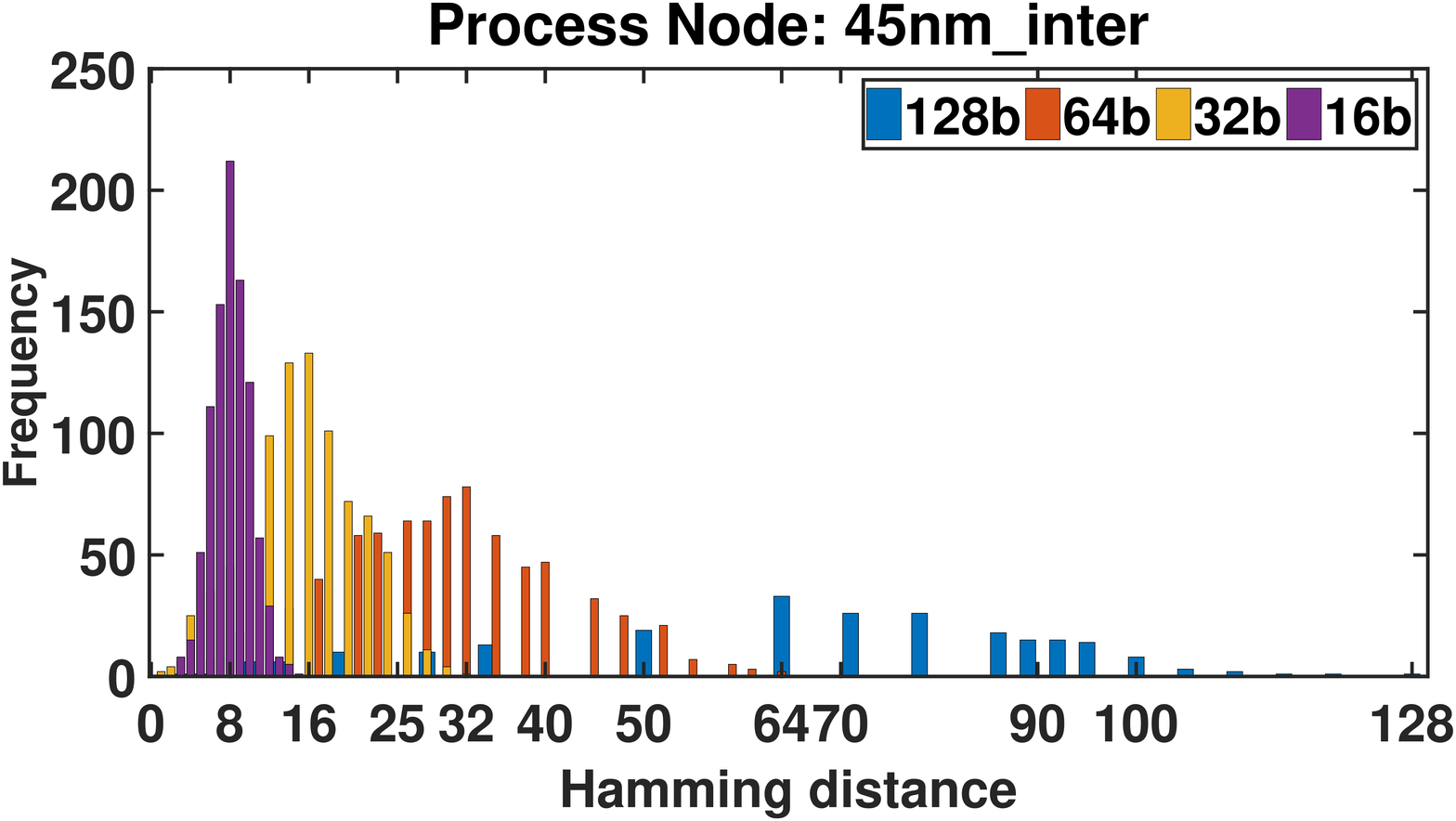} & 
\includegraphics[width=\textwidth]{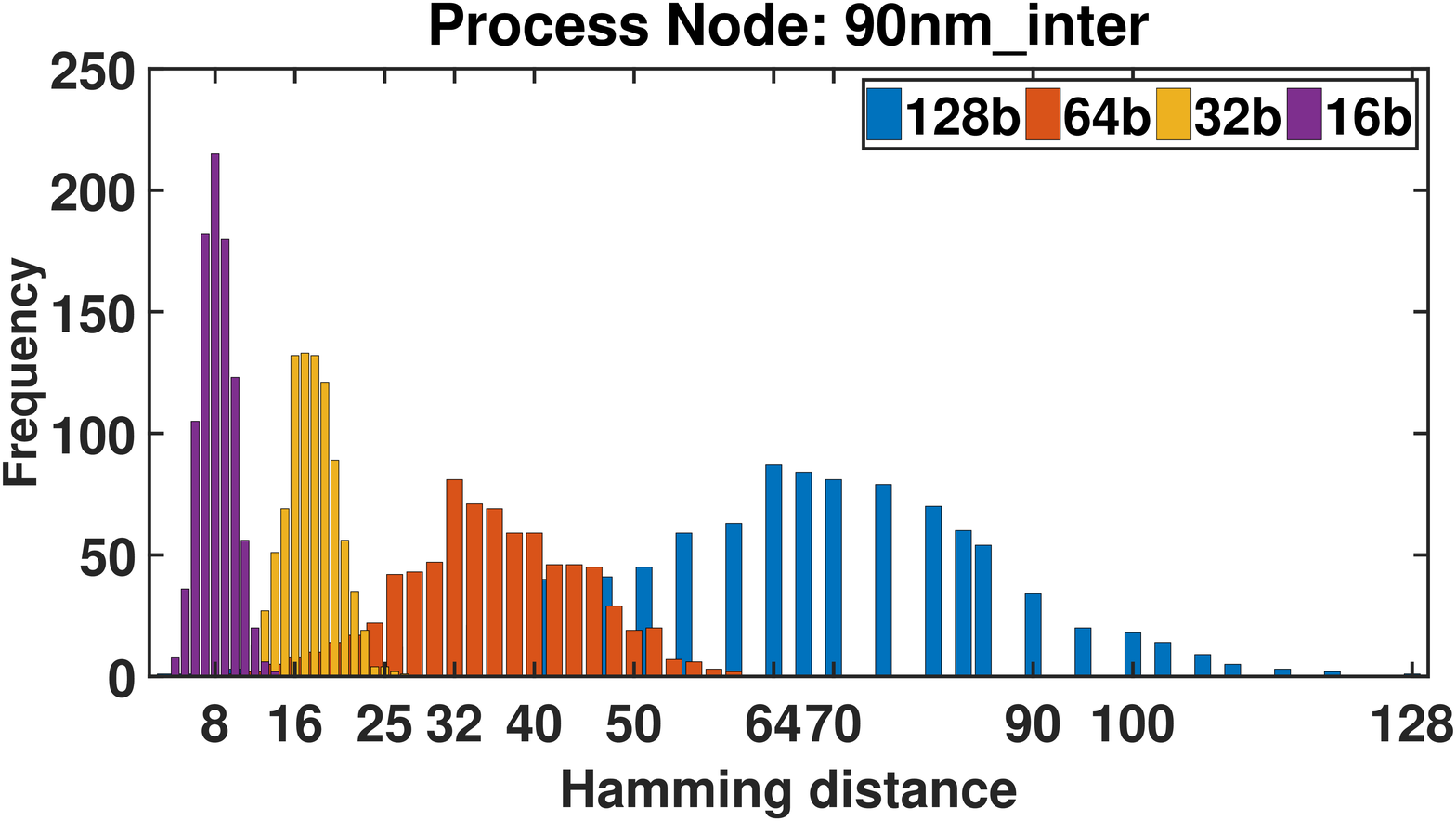} \\ 
{\Huge (a)} & {\Huge(b)} & {\Huge(c)} \\
\includegraphics[width=\textwidth]{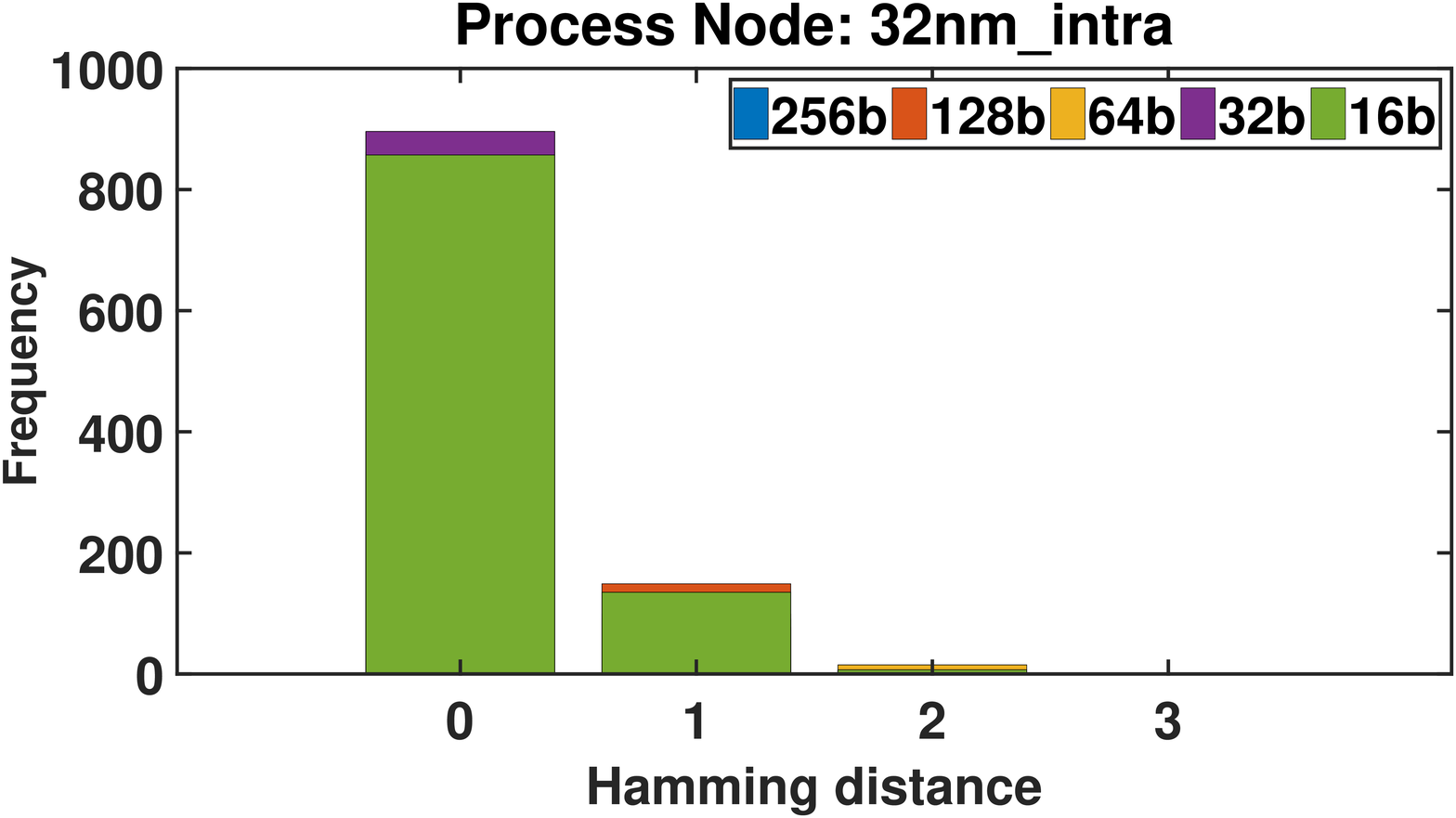} & 
\includegraphics[width=\textwidth]{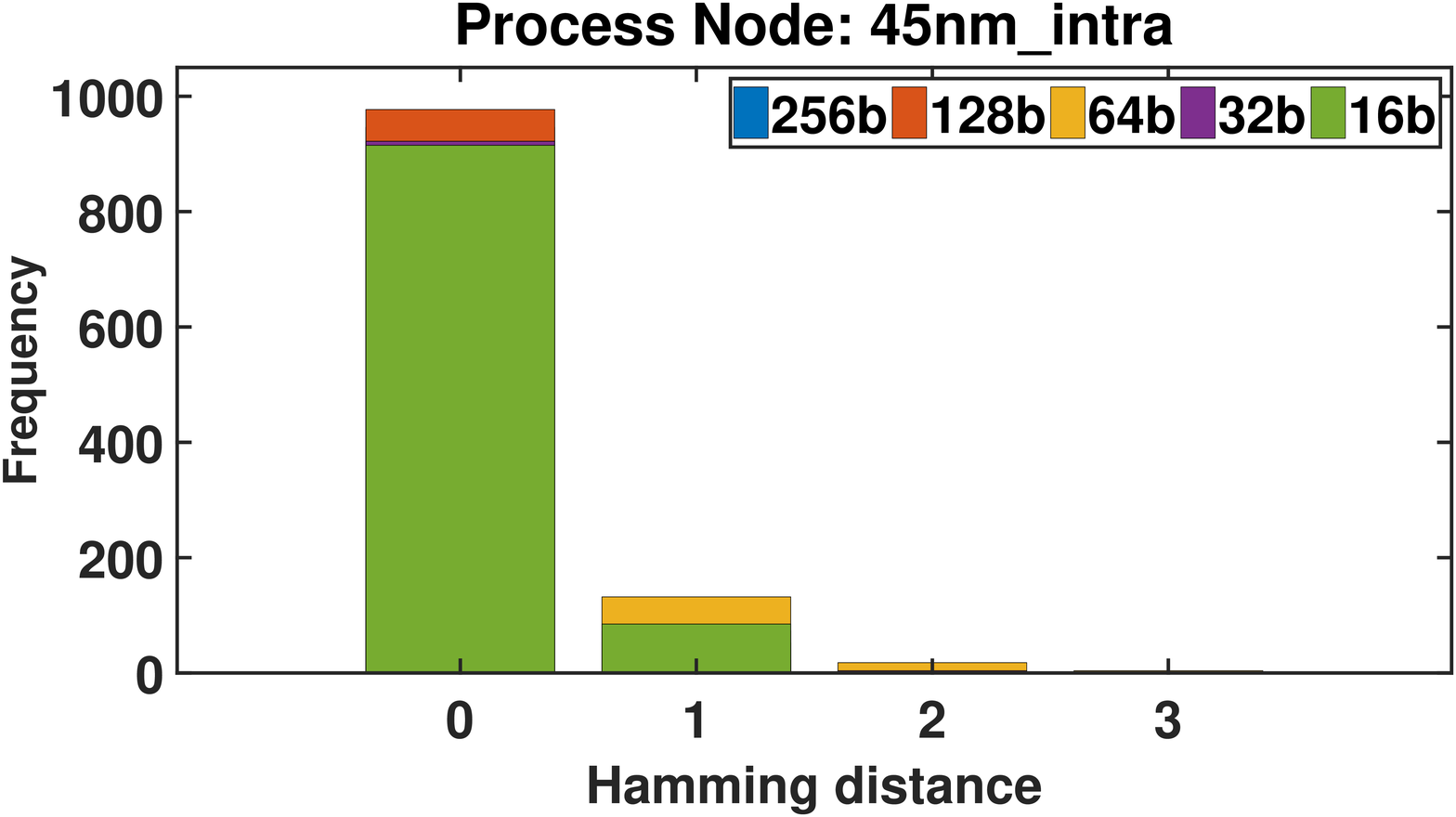} & 
\includegraphics[width=\textwidth]{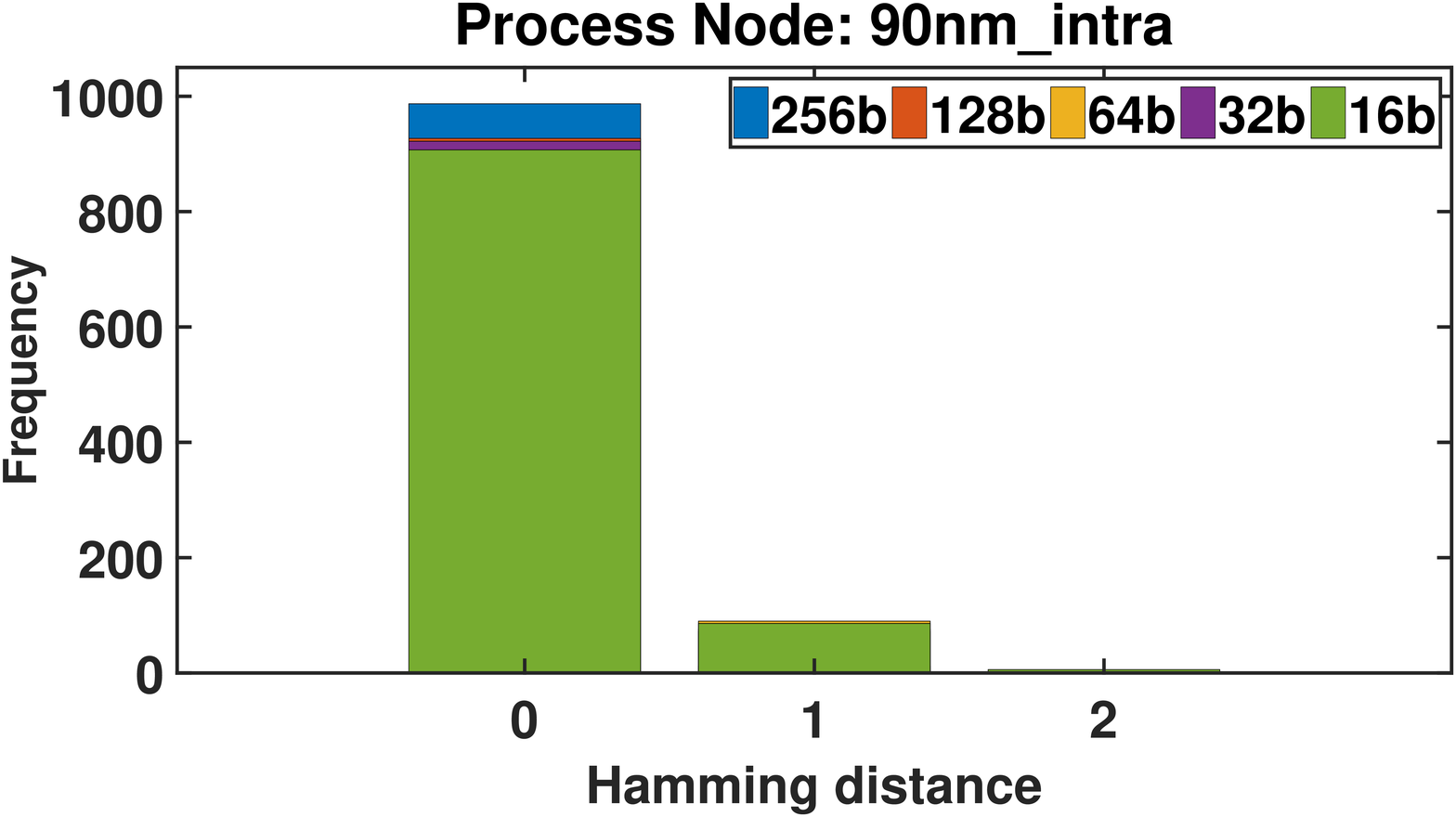} \\ 
{\Huge (d)} & {\Huge(e)} & {\Huge(f)} \\
\end{tabular}}
\vspace*{-1ex}
\caption{Overlay Histogram of inter-chip HD (a-c) and intra-chip HD (d-f) for four key lengths  in the three process nodes. }
\label{fig:uniqueness}
\vspace*{-3ex}
\end{figure*}




\begin{figure*}[h]
\centering
\resizebox{\textwidth}{!}{
\begin{tabular}{ccc}
\includegraphics[width=\textwidth]{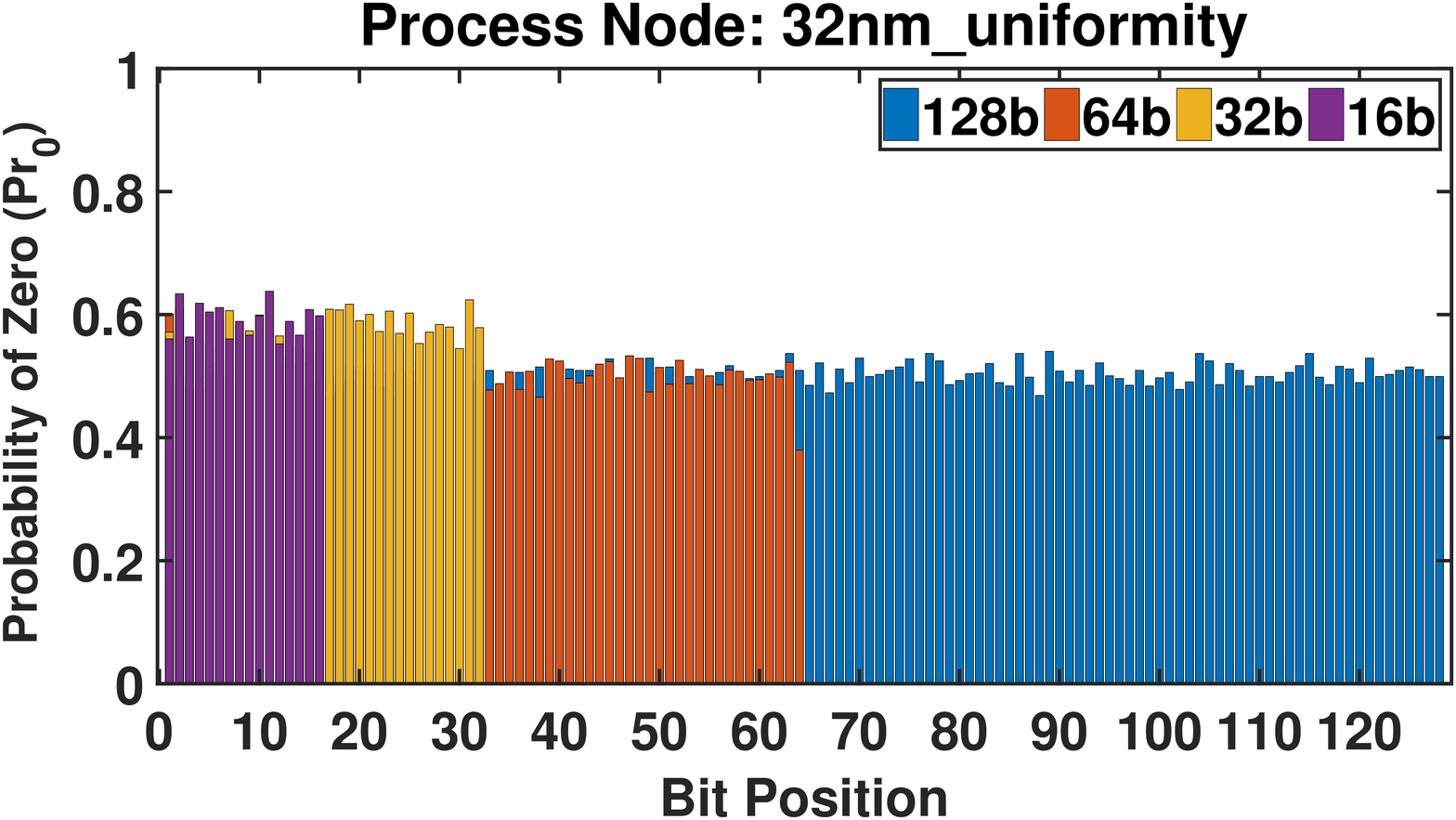} & 
\includegraphics[width=\textwidth]{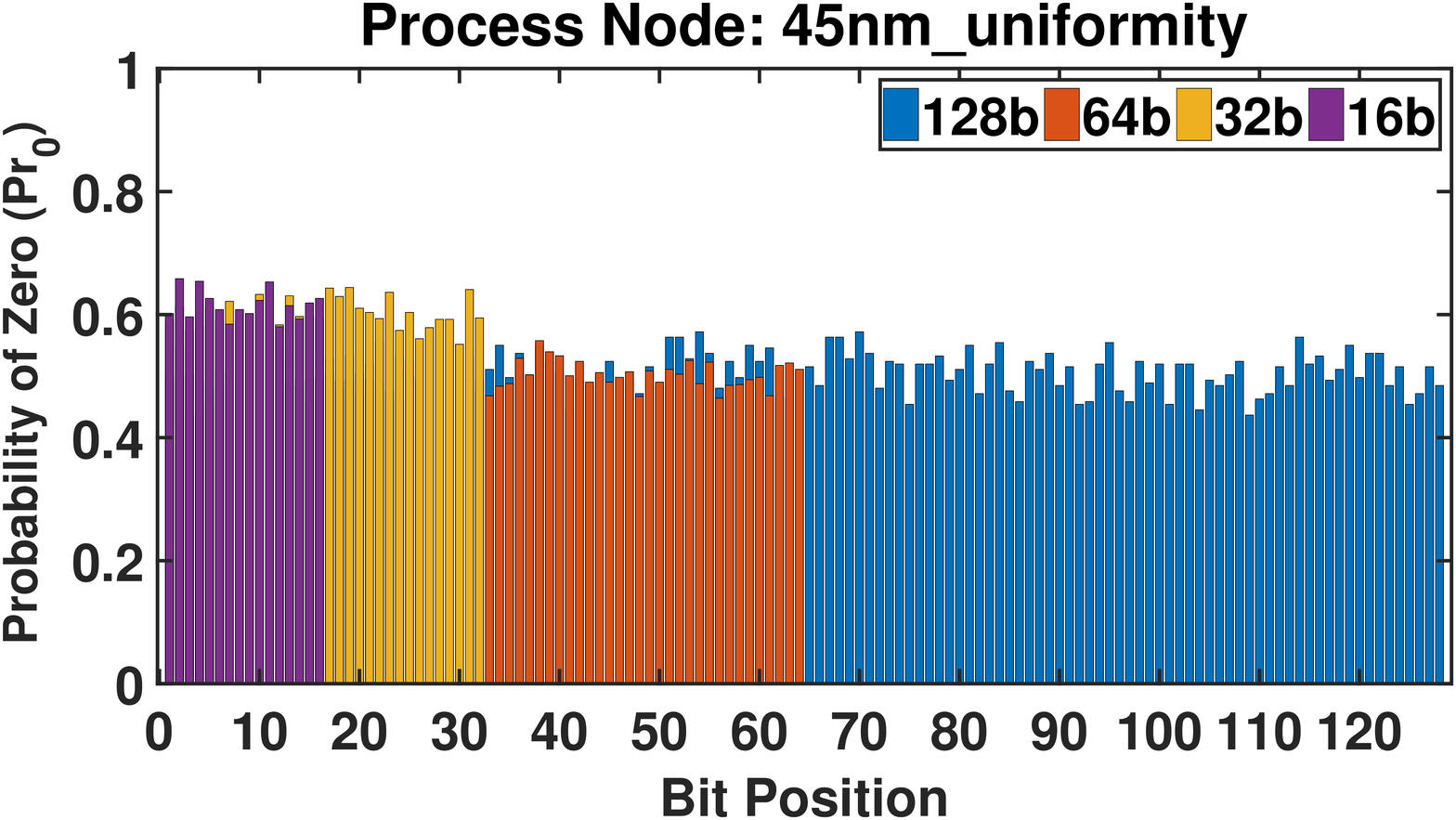} & 
\includegraphics[width=\textwidth]{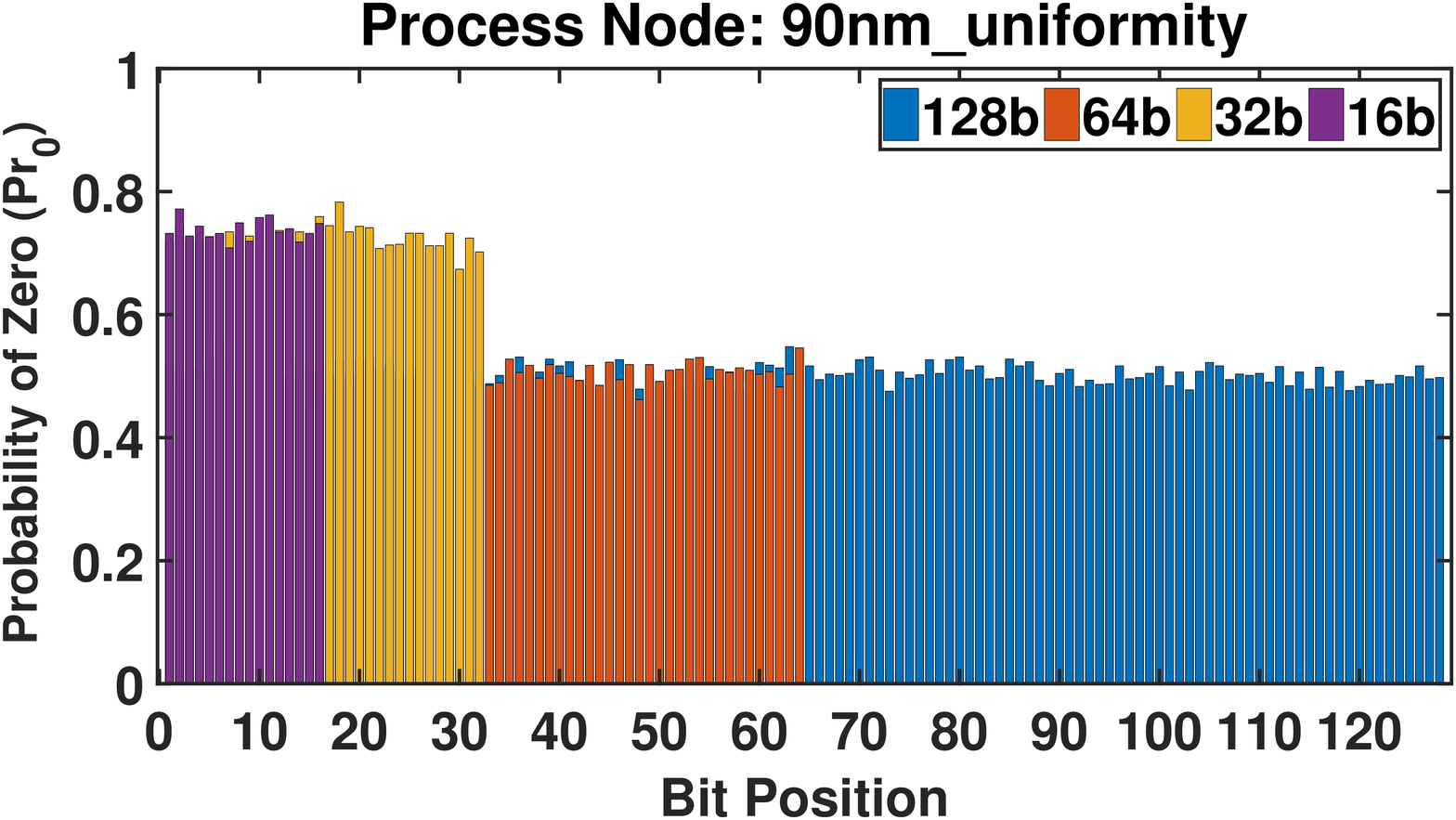} \\ 
{\Huge (a)} & {\Huge(b)} & {\Huge(c)} \\
\end{tabular}}
\vspace*{-1ex}
\caption{Probability of zero in each bit position  for four (16-,
32-, 64-, and 128-bit) key lengths in three process nodes.}
\label{fig:uniformity}
\vspace*{-3ex}
\end{figure*}


\begin{figure*}[h]
\centering
\resizebox{\textwidth}{!}{
\begin{tabular}{ccc}
\includegraphics[width=\textwidth]{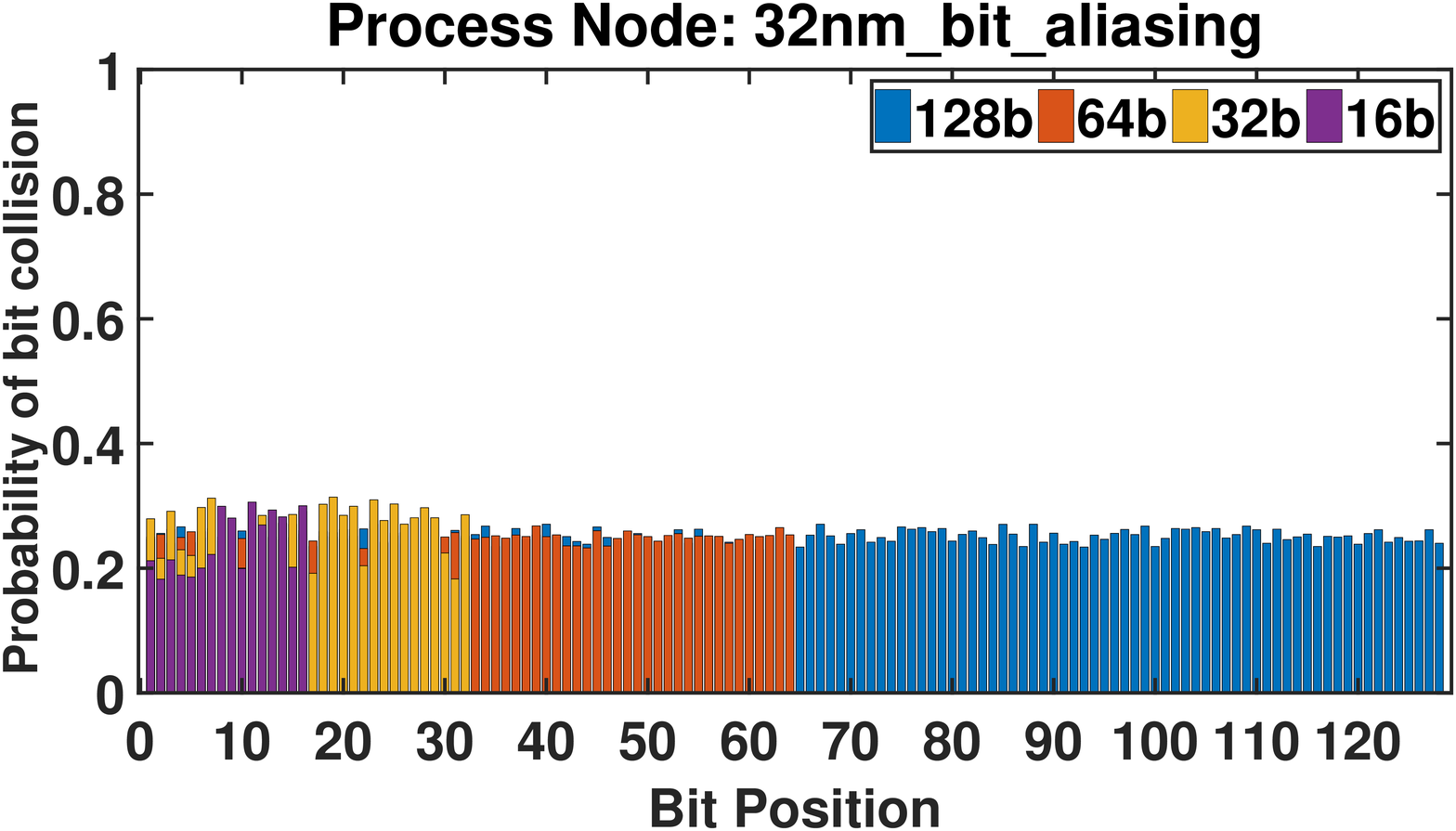} & 
\includegraphics[width=\textwidth]{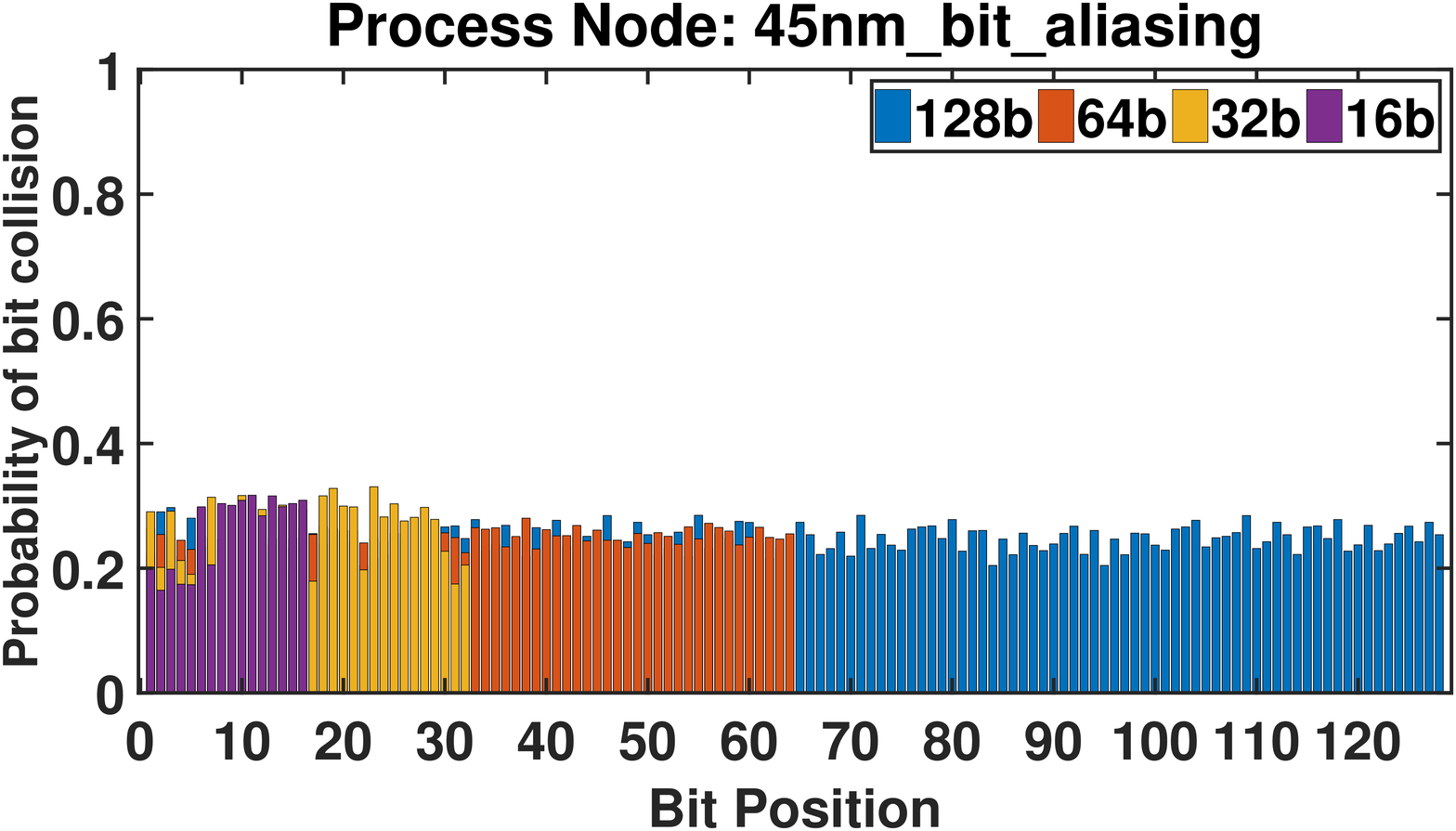} & 
\includegraphics[width=\textwidth]{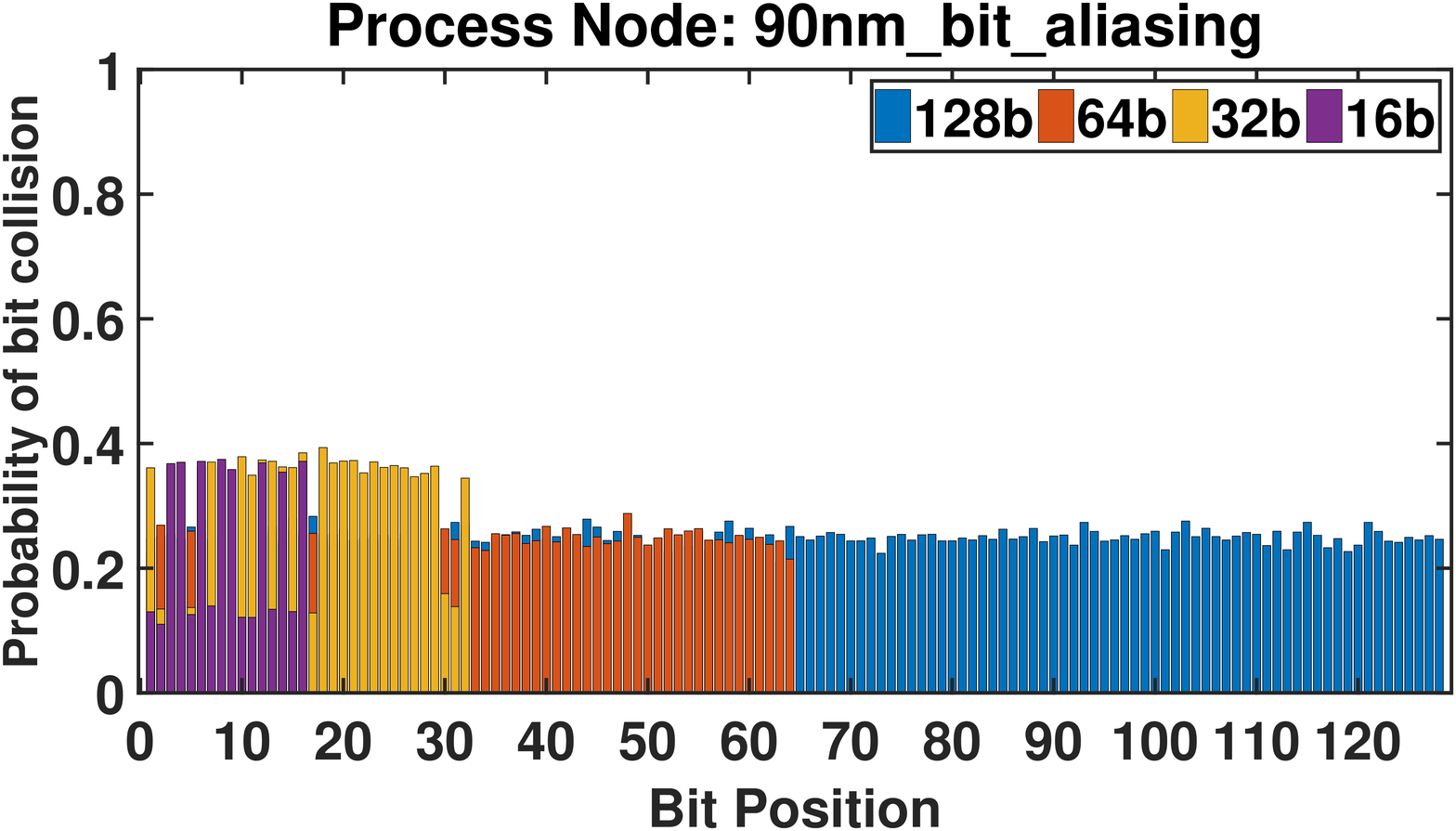} \\ 
{\Huge (a)} & {\Huge(b)} & {\Huge(c)} \\
\end{tabular}}
\vspace*{-1ex}
\caption{Average probability of collision in each bit position. }
\label{fig:collision}
\vspace*{-2ex}
\end{figure*}

We perform Monte Carlo (MC) simulations of SR-FF PUF  at SPICE level using Synopsys HSPICE for three CMOS processes (90nm, 45nm, and 32nm) \cite{PTM}. MC can perform device variability analysis within six-sigma limit, hence the Challenges-Response Pairs (CRPs) collected using MC is comparable to CRPs from manufactured ICs. We simulate the PUF structure for 1000 iterations, analogous to 1000 different dies on a 300mm wafer at nominal voltage (1V).
Several works \cite{1561249,4261134,6587277} in the literature have validated PUF design through SPICE level simulations. We then evaluate PUF responses according to the parameters  proposed by Maiti {\em et al.} \cite{cryptoeprint:2011:657}. 
Although process variations impact the channel length, we maintain length variability within (intra-die) 15\% and across (inter-die) 33\% of nominal value  to generate CRPs \cite{839819}. We also report the performance overhead of physical synthesis for five RTL designs \cite{OpenCores} with centroid architecture.

\textbf{Uniqueness}: Uniqueness provides the measurement of inter-chip variation. We can measure the uniqueness
by calculating Hamming Distance (HD) of two pair-wise dies. Ideally, two dies (chips) show a distinguishable response (HD $\sim$ 50\%) to a common challenge. Fig. \ref{fig:uniqueness}(a-c) shows inter-chip HD of four different key sequences. For all keys, we made two thousand comparisons to verify 
uniqueness. One can see that the average HD for all key-lengths are close to 49\%.

\begin{table}
\caption{Specifications of wireload model in three design corners (cap. unit, fF and res. unit, K\si{\ohm})}
\vspace{-1ex}
\label{tab:wireloadrc}
\begin{center}
\resizebox{\columnwidth}{!}{\begin{tabular}{|c|c|c|c|c|c|c|c} 
\toprule
\multirow{2}{*}{Wireload Model} & \multicolumn{2}{c|}{Best} & \multicolumn{2}{c|}{Typical} & \multicolumn{2}{c|}{Worst} \\ \cline{2-7}
 & Cap.  & Res. & Cap. & Res. & Cap. & Res. \\  \midrule
8000 & 0.00028 &	1.42E-03& 0.000312	&1.57E-03 &0.000343	& 1.73E-03 \\
16000&	0.000512&	1.15E-03& 0.000569&	1.28E-03& 0.000625& 	1.41E-03 \\
35000 &	0.000243&	1.07E-03 &0.00027	&1.19E-03 &0.000297	& 1.31E-03 \\
70000&	0.000128&	9.00E-03& 0.000143&	1.00E-02& 0.000157&	1.10E-02 \\ \midrule
\end{tabular}}
\end{center}
\vspace{-3ex}
\end{table}

\begin{table}
\caption{Wire specifications across all wireload models in three design corners}
\vspace{-1ex}
\label{tab:wirespec}
\begin{center}
\begin{tabular}{c c} 
\toprule
Wire width & (0.45, 0.9, 1.35, 1.8, 2.25) \\ \midrule
\multirow{2}{*}{Wire Spacing} & (0.45, 0.9, 1.35, 1.8, 2.25, 2.7, 3.15, 
 3.6, \\ & 4.05, 4.5, 4.95, 5.4, 5.85, 6.3, 6.75, 7.2) \\
 \midrule
\end{tabular}
\end{center}
\vspace{-7ex}
\end{table}

\begin{table*}
\begin{center}
\caption{Overhead comparison of design attributes in SR-FF PUF with centroid architecture}
\vspace{-1ex}
\label{tab:overhead}
\resizebox{\textwidth}{!}{\begin{tabular}{|c|c|c|c|c|c|c|c|c|c|c|c|c|c|c|c|c|c|c|c|c|c|c|}
\hline
\multirow{2}{*}{Design} & \multirow{2}{*}{No. of bits} & \multicolumn{3}{c|}{Best-case}  & \multicolumn{3}{c|}{Typical-case} & \multicolumn{3}{c|}{Worst-case}   \\ \cline{3-11}
& & Area (\%) & Power (\%) & Delay (\%) & Area (\%) & Power (\%) & Delay (\%) & Area (\%) & Power (\%) & Delay (\%)  \\ \midrule

AES128 &  1072 & 0.009 & 4.065&	2.622& 0.017 & 1.301	 &	3.836 & 0.320 &	0.473 &	6.671  \\ \midrule
DES & 1827 & 0.022 &0.963 &0.824 &0.037 &0.39 &1.923 &0.604 &0.058 &3.968\\ \midrule
Triple DES & 2083 &  0.010 &0.698 &0.781 &0.035 & 0.510 &1.858 &0.711 &0.067 &3.095\\ \midrule
8-bit uP & 386 &  0.584 & 4.884 & 0 & 2.51 & 2.658 & 0.175 & 4.185 & 0.021 & 6.096 \\ \midrule
Cannyedge Detector & 2027 &  0.109 &	1.587 &  1.354		& 2.487 & 	0.164 &	5.812  & 3.585&	0.101 &	6.676 \\ \midrule
Average & & 0.146 & 2.439 & 1.116 &  1.017 & 1.004 & 2.720 & 1.881 & 0.144 & 5.301 \\ \midrule

\end{tabular}}
\end{center}
\vspace*{-7ex}
\end{table*}

\textbf{Reliability}: We can measure the reliability from Bit Error Rate (BER) of PUF responses for intra-chip variation. Ideally, a PUF should maintain the same response
(100\% reliable or 0\% variation) on different environmental variations (supply voltage, temperature) under the same challenge.  Fig. \ref{fig:uniqueness}(d-f) shows the intra-die HD for five key length in three process nodes different temperature ($0^{o}$C to $80^{o}$C).  The reliability (HD = 0) for 16-, 32-, 64-, and 128-bit registers are 92.3\%, 92.2\%, 90.7\%, and 92.7\% respectively.

\textbf{Uniformity/Randomness}: Uniformity measures the ability of a PUF to generate uncorrelated `0's and `1's in the response. Ideally, PUF should generate `0's and
`1's with equal probability in a response. This ensures the resilience of guessing PUF response from a known challenge. The probability of zero is bound within 0.5 and 0.7 for four different key lengths in Fig. \ref{fig:uniformity}. Although the ideal value of uniform probability should be 0.5, variability in gate delay due to process variability impacts the even distribution of `0's and `1's.

\textbf{Bit aliasing/Response collision}: To evaluate the bit aliasing, we use the same set of responses used in uniqueness. We see the average probability of collision less than 30\% as shown in Fig. \ref{fig:collision}. As the reference response is chosen randomly and compared to the rest of the responses, an adversary can guess, on average, less than 30\% of the correct responses. Hence, the generated responses are resistant to the key-guessing attack.


\textbf{Physical Synthesis Analysis}: We report the physical synthesis results of designs from OpenCores \cite{OpenCores}. We perform the logic synthesis with Synopsys Design Compiler and  the layout (floor planning, placement, and routing)  of the mapped netlist  using Synopsys IC Compiler. We evaluate the area, power, and timing overhead for SAED 90nm technology in three design corners, namely, best, typical and worst.

Table \ref{tab:wireloadrc} lists the required resistance and capacitance (routing and parasitic) values during cell characterization for achieveing  metastable state. The inter-transistor routing across all wireload models are presented in Table \ref{tab:wirespec}. The capacitance values include both routing and parasitic capacitance. We  vary input voltages (0.7V-1.32V) with (\texttt{on\_chip\_variation}) enabled during synthesis. It confirms that the responses are not biased  towards a particular input voltage value and adversary can not further tamper the device responses with aggressive supply voltage changes.
We maintain a 4by4 grid across all designs to implement centroid architecture and distribute it randomly. Depending on the dimension of the grid, the total number of the grid would grow or shrink. Following that, we report the overhead after physical synthesis in Table \ref{tab:overhead}. The number of bits in Table \ref{tab:overhead} represent the possible key length of design. Across different wireload models of a particular design corner, we observe more delay and power variations  due to variable resistance and capacitance. For 8-bit uP, the centroid architecture  is adjacent to high-activity logic, hence we see increased PPA overhead. In the remaining designs, best-case minimizes the area and delay overhead and during worst-case, we see a reduction in power overhead.

\section{Conclusion}
\label{sec:conclusion}

In this work, we have proposed to use the existing SR flip-flop in the design to quantify its race condition for PUF implementation. We embed a centroid architecture with SR FFs so that PUF responses conform to local transistor variations only. The generated responses exhibit better uniqueness, randomness, reliability and reduced bit-aliasing compared to other metastability-based PUFs. For future work, we would evaluate the uniqueness of SR-FF PUF responses from transient noise based simulation and  the resilience  against adversarial machine learning attack.

\bibliographystyle{unsrt}
\scriptsize{
\bibliography{bib/main.bib}

\begin{thebibliography}{10}

\bibitem{8031550}
M.~Fyrbiak \emph{et. al.}
\newblock {Hardware reverse engineering: Overview and open challenges}.
\newblock In {\em 2017 IEEE IVSW}, pages 88--94, July 2017.

\bibitem{1561249}
D.~Lim \emph{et. al.}
\newblock {Extracting secret keys from integrated circuits}.
\newblock {\em IEEE TVLSI}, 13(10):1200--1205, Oct 2005.

\bibitem{HABIB201792}
B.~Habib et~al.
\newblock Implementation of efficient sr-latch puf on fpga and soc devices.
\newblock {\em Microprocessors and Microsystems}, 53:92 -- 105, 2017.

\bibitem{7881790}
A.~Ardakani and S.~B. Shokouhi.
\newblock {A secure and area-efficient FPGA-based SR-Latch PUF}.
\newblock In {\em 2016 IST}, pages 94--99, Sept 2016.

\bibitem{8351052}
S.~Wang \emph{et al.}
\newblock {Register PUF with No Power-Up Restrictions}.
\newblock In {\em 2018 IEEE ISCAS}, pages 1--5, May 2018.

\bibitem{BossuetNCF14}
L.~Bossuet \emph{ et. al.}
\newblock {A {PUF} Based on a Transient Effect Ring Oscillator and Insensitive
  to Locking Phenomenon}.
\newblock {\em {IEEE} TETC}, 2(1):30--36, 2014.

\bibitem{7428066}
S.~S.~Zalivaka \emph{et. al.}
\newblock {Multi-valued Arbiters for quality enhancement of PUF responses on
  FPGA implementation}.
\newblock In {\em 2016 ASP-DAC}, pages 533--538, Jan 2016.

\bibitem{8357321}
S.~A. {Islam} and S.~{Katkoori}.
\newblock {High-level synthesis of key based obfuscated RTL datapaths}.
\newblock In {\em 2018 19th International Symposium on Quality Electronic
  Design (ISQED)}, pages 407--412, March 2018.

\bibitem{8607170}
S.~A.~{Islam} \emph{et. al.}
\newblock {Empirical Word-Level Analysis of Arithmetic Module Architectures for
  Hardware Trojan Susceptibility}.
\newblock In {\em 2018 Asian Hardware Oriented Security and Trust Symposium
  (AsianHOST)}, pages 109--114, Dec 2018.

\bibitem{8491861}
J.~Danger \emph{et. al.}
\newblock {Analysis of Mixed PUF-TRNG Circuit Based on SR-Latches in FD-SOI
  Technology}.
\newblock In {\em 2018 DSD}, pages 508--515, Aug 2018.

\bibitem{4443209}
Y.~Su, J.~Holleman, and B.~P. Otis.
\newblock {A Digital 1.6 pJ/bit Chip Identification Circuit Using Process
  Variations}.
\newblock {\em IEEE Journal of Solid-State Circuits}, 43(1):69--77, Jan 2008.

\bibitem{PTM}
{Predictive Technology Model}.
\newblock \url{http://ptm.asu.edu/}.

\bibitem{4261134}
G.~E. {Suh} and S.~{Devadas}.
\newblock {Physical Unclonable Functions for Device Authentication and Secret
  Key Generation}.
\newblock In {\em 2007 44th ACM/IEEE DAC}, pages 9--14, June 2007.

\bibitem{6587277}
U.~{Rührmair} \emph{et. al.}
\newblock {PUF Modeling Attacks on Simulated and Silicon Data}.
\newblock {\em IEEE TIFS}, 8(11):1876--1891, Nov 2013.

\bibitem{cryptoeprint:2011:657}
A.~Maiti \emph{et. al.}
\newblock {A Systematic Method to Evaluate and Compare the Performance of
  Physical Unclonable Functions}, 2011.

\bibitem{839819}
S.~Nassif.
\newblock {Delay variability: sources, impacts and trends}.
\newblock In {\em 2000 IEEE ISSCC}, pages 368--369, Feb 2000.

\bibitem{OpenCores}
{OpenCores}.
\newblock \url{https://opencores.org/}.

\end{thebibliography}
}
\end{document}